\documentclass{elsart3p}
\usepackage{graphicx,natbib,amssymb}
\journal{Physics Letters A}
\makeatletter
\def\elsartstyle{%
    \def\normalsize{\@setfontsize\normalsize\@xiipt{14.5}}
    \def\small{\@setfontsize\small\@xipt{13.6}}
    \let\footnotesize=\small
    \def\large{\@setfontsize\large\@xivpt{18}}
    \def\Large{\@setfontsize\Large\@xviipt{22}}
    \skip\@mpfootins = 18\p@ \@plus 2\p@
    \normalsize
}
\@ifundefined{square}{}{\let\Box\square}
\makeatother

\begin{document}

\begin{frontmatter}
\title{Theory on Plasmon Modes of the Cell Membranes}

\author[IPE]{T. T. Nhan},
\author[IPE]{N. T. Nhan},
\author[IPE,APCTP]{N. V. Thanh},
\corauth[cor]{Email address: nvthanh@iop.vast.ac.vn}
\author[IPE]{N. A. Viet}
\address[IPE]{Institute of Physics and Electronics,\\ P. O. Box 429, Boho, Hanoi 10000, Vietnam.}
\address[APCTP]{APCTP, Hogil Kim Memorial Building 5th floor,\\ POSTECH, Hyoja-dong, Namgu, Pohang 790-784, Korea.}
\begin{abstract}
Considering the plasmon oscillation of each layer of the cell membranes as a quasi-particle, we introduce a simple model for the membrane collective charge excitations, take into account the surface effective potential of the plasmon-plasmon interaction between two layers. By using the useful Bogoliubov transformation method, we easily obtained the expressions of the frequencies of plasmon oscillations as a function of wave-number $k$ and membrane thickness~$d$, magnitude of these frequencies is in the order of $\sqrt{kd}$. Our results are in good agreement with ones obtained by E.~Manousakis \cite{Manousakis}.
\end{abstract}

\end{frontmatter}

\section{Introduction}
\label{intro}

The biology and physics \cite{Safran} communities have pursued overlapping studies of membranes over the past few decades, with the natural segmentation into more specific (biological) and general (physics) studies. The physicists are often entranced by the variety of topological and phase behaviours possible in membranes. However, with new experimental techniques that include better visualization of dynamical processes in cells, there are increasingly more opportunities for interdisciplinary cross-fertilization.

Plasmon modes have been found \cite{Grimes} in a system of electrons on the surface of liquid helium, that play a fundamental role in understanding the properties of a metal in condensed matter. Plasmons have been investigated both theoretically \cite{Manousakis,Ritchie,Ferrell,FStern,Fetter} and experimentally \cite{Geballe} in quasi $2D$ systems such as surfaces or interfaces of metals. 

Lipid bilayers are among the most important "construction materials" for the cell \cite{Alberts}. A bilayer membrane constitutes a flexible barrier that separates the interior and exterior of a cell, encapsulates the nucleus and can perform a number of roles such as acting as a functional host for protein production. Membranes appear in flat (plasma membrane), spherical (vesicular transport), tubular (transport) or tortuous
and possibly bicontinuous (Endoplasmic Reticulum, ER and Golgi apparatus) forms in the cell, depending on function and composition.
The surface of a lipid bilayer with the surrounding counter-ion system is analogous to that of electrons on helium surface. The surface of a cell membrane in aqueous environment becomes negatively charged \cite{Hille}. It is well known \cite{Grahame,Stern} that a diffuse double layer of counter-ions from the solution screens the negative charge, thus, a thin two-dimensional $(2D)$ layer of mobile cations is accumulated adjacent to the extracellular and intra-cellular membrane surfaces. In addition, because of different permeabilities of the cell membrane channels to various cations, the inside of the cell is kept at a negative potential relative to the outside. The gating mechanism of ion channels along the cell membrane, that are believed to open using voltage sensing gates, is an open problem in cell biology \cite{Hille,Ahern,Lodish}.

Recently, the collective charge response of the mobile counter-ion charge of the interior and exterior surface layers to an external electric field perturbation have been studied by E.~Manousakis \cite{Manousakis}. Following a complicated approach, one have to solve the Poisson equation for the perturbation field due to the external charge density and current fluctuations, one obtained the expressions of the frequencies of plasmon oscillations of each layer. It is shown that there are longitudinal charge collective modes with wave vectors parallel to the surface and the two layers are coupled and fluctuate together as a whole in the limit where the wavelength is much longer than the membrane thickness $d$.

In this work we introduce a simple model for the membrane collective charge excitations. The formulae of the frequencies of plasmon oscillations are obtained by using the Bogoliubov transformation method.

\section{Model}
\label{model}

\begin{figure}[t]
\begin{center}
\includegraphics*[width=5cm]{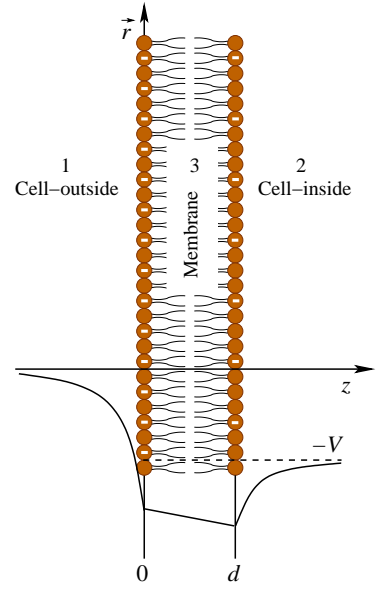}
\end{center}
\caption{The model of lipid bilayer cell membrane of width $d$.}
\label{fig:model}
\end{figure}

The negatively charged exterior and interior surfaces at biological membranes of cells is shown in Fig.~\ref{fig:model} where the density of charged phospholipids heads is higher on the exterior membrane surface than interior one. Diffuse double layers of counter-ions, such as H$^+$, Na$^+$, K$^+$, Ca$^{++}$..., form on each membrane side. Due to the difference membrane permeability for these cations, there is a voltage $-V$ between the cell interior relative to the exterior. 

In the thermal equilibrium the counter-ion charge density varies as we move away from the membrane surface as shown, within in the so-called Gouy-Chapman (GC) length 
$$\lambda^i_{GC}=\frac{k_BT\varepsilon_i}{2\pi\sigma_iq_i}$$
where $i=1,\ 2$, for each membrane side. $\varepsilon_i$ and $q_i=Z_ie$ are the dielectric constants and the counter-ion charges of the outside and inside of the cell with $Z_i$ being the ionic valence. At the room temperature $T=300K$, ones obtained $\lambda_{GC}^{-1}\sim 2.5 nm^{-1}$. For charge fluctuations propagating with wave vectors $\mathbf k$ parallel to the surfaces and magnitude $k \ll \lambda_{GC}^{-1}$, these layers can be regarded as two-dimensional \cite{Manousakis}.

We suppose the plasmons of each layer as a quasi particle with a fluctuation frequency $\omega (\mathbf k)$
\begin{equation}
\omega_i(\mathbf k)=Z_i\omega_P+s_ik,
\label{eq:01}
\end{equation}
in which, the notations $\omega_P$ and $s_i$ is the $2D$ plasmon frequency and the velocity of density fluctuations, respectively:
\begin{eqnarray}
\label{eq:02}
\omega_P&=&\sqrt{\frac{2\pi e^2n_0k}{m_i\varepsilon_i}},\\
\label{eq:03}
s_i&=&\sqrt{\frac{2k_BT}{m_i}},
\end{eqnarray}
with $n_0\approx 1nm^{-2}$ being the surface density taken from experiments, and $m_i$ are the counter-ion-effective masses. The coupling coefficients between the plasmon of two layers play the role of the surface effective potentials which can be taken in the form
\begin{equation}
\varphi_i=Z_1Z_2\omega_Pe^{-kd}.
\label{eq:04}
\end{equation}
The Hamiltonian for the plasmons in cell membranes is defined by
\begin{eqnarray}
{\cal H}_m&=&\sum_k\left[\omega_1(\mathbf k)a^+(\mathbf k)a(\mathbf k)+\omega_2(\mathbf k)b^+(\mathbf k)b(\mathbf k)\right.\nonumber\\
&&\left.+\varphi_1(\mathbf k)b^+(\mathbf k)a(\mathbf k)+\varphi_2(\mathbf k)a^+(\mathbf k)b(\mathbf k)\right],
\label{eq:05}
\end{eqnarray}
where $a^+$ ($a$) and $b^+$ ($b$) are the creation (annihilation) operators of the plasmons on exterior and interior membrane sides, respectively. These operators obey the commutation relations:
\begin{eqnarray}
&&\left[a,a^+\right]=\left[b,b^+\right]=1,\nonumber\\ 
\label{eq:comp}
&&\left[a,a\right]=\left[a^+,a^+ \right]=\left[b,b\right]=\left[b^+,b^+ \right]=0,\\
&&\left[a,b\right]=\left[a,b^+ \right]=\left[a^+,b\right]=\left[a^+,b^+ \right]=\ldots=0.\nonumber
\end{eqnarray}

Using the Bogoliubov transformation, we can rewrite the Hamiltonian (\ref{eq:05}) in a diagonalized form
\begin{eqnarray}
{\cal H}_m&=&\sum_k\left[\Omega_+(\mathbf k)\alpha^+(\mathbf k)\alpha(\mathbf k)\right.\nonumber\\
&&+\left.\Omega_-(\mathbf k)\beta^+(\mathbf k)\beta(\mathbf k)\right],
\label{eq:06}
\end{eqnarray}
where  $\alpha^+$ ($\alpha$) and $\beta^+$ ($\beta$) are the creation (annihilation) operators of new quasi-particles that obey commutation relations (\ref{eq:comp}). The Eigen-frequencies of the membrane collective charge oscillations can be obtained by solving the determinant equation
\begin{equation}
\left|
  \begin{array}{cc}
   \omega_1(\mathbf k)-\Omega_+(\mathbf k)& \varphi_1(\mathbf k) \\
    \varphi_2(\mathbf k) & \omega_2(\mathbf k)-\Omega_+(\mathbf k)\\
  \end{array}
\right|=0,
\label{eq:07}
\end{equation}
then we have
\begin{eqnarray}
\Omega_\pm(\mathbf k)&=&\frac{\omega_1(\mathbf k)+\omega_2(\mathbf k)}{2}\nonumber\\
&\pm&\frac{1}{2}\sqrt{\left[\omega_1(\mathbf k)-\omega_2(\mathbf k)\right]^2+4\varphi_1(\mathbf k)\varphi_2(\mathbf k)}.
\label{eq:08}
\end{eqnarray}
The frequencies denoted by $\Omega_+$ ($\Omega_-$) correspond to in-phase (out-of-phase) oscillations of both layers
together. It means that the frequencies of plasmon oscillations attain maximum value at $\Omega(\mathbf k)=\Omega_+(\mathbf k)$ and minimum value at $\Omega(\mathbf k)=\Omega_-(\mathbf k)$.

\section{Results and Discussion}
\label{model}

We consider a simplest case where all the parameters of both layers are taken to be the same i.e., $\omega_1(\mathbf k)=\omega_2(\mathbf k)$ and $\varphi_1(\mathbf k)=\varphi_2(\mathbf k)$. The frequencies are given in units of 
\begin{equation}
\omega_0=Z\sqrt{\frac{2\pi e^2n_0}{md\varepsilon}}.
\end{equation}
The expression (\ref{eq:08}) becomes
\begin{equation}
\frac{\Omega_\pm(\mathbf k)}{\omega_0}=\sqrt{kd}+k\sqrt{dd_0}\pm\sqrt{kd}\,e^{-kd},
\label{eq:09}
\end{equation}
in which 
$$d_0=\frac{k_BT\varepsilon}{\pi e^2 n_0}\simeq 5\AA$$
is taken from the experiments. We see that the magnitude of the frequencies is the in order of $\sqrt{kd}$. These results are in quite good agreement with the obtained results in Ref.~\cite{Manousakis}.

\begin{figure}[b]
\begin{center}
\includegraphics*[width=7.5cm]{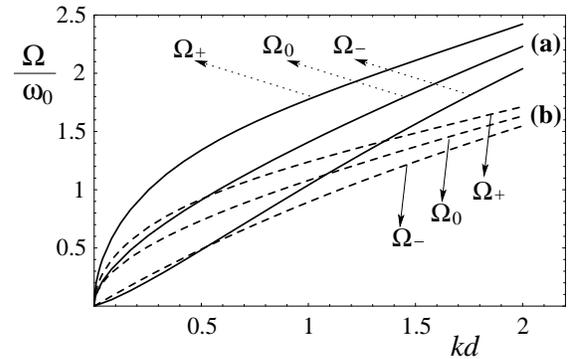}
\end{center}
\caption{Eigen-frequencies of the membrane, with $d\simeq 30\AA$. The dashed curves are taken from Ref.~\cite{Manousakis} for comparison.}
\label{fig:freq}
\end{figure}

In Fig.~\ref{fig:freq} we present the Eigen-frequencies of the membrane collective charge oscillations versus $kd$, where  the wave number $k$ is given in units of $1/d$. The solid curves correspond to our calculations (a) and the dashed curves are included from obtained results of E.~Manousakis (b) for comparison. The middle curve of part (a) or (b) show the frequencies of each independent layer, i. e., there is no coupling between the layers, then the frequencies can be expressed as 
$$\frac{\Omega_0(\mathbf k)}{\omega_0}=\sqrt{kd}+k\sqrt{dd_0}.$$

It is straightforward to show that in the long-wavelength limit ($kd \ll 1$)
\begin{eqnarray}
\frac{\Omega_+(\mathbf k)}{\omega_0}&\approx& 2\sqrt{kd}-\sqrt{(kd)^3}+k\sqrt{dd_0},\\
\frac{\Omega_-(\mathbf k)}{\omega_0}&\approx&\sqrt{(kd)^3}+k\sqrt{dd_0}\,.
\end{eqnarray}

\begin{figure}[h]
\begin{center}
\includegraphics*[width=7.5cm]{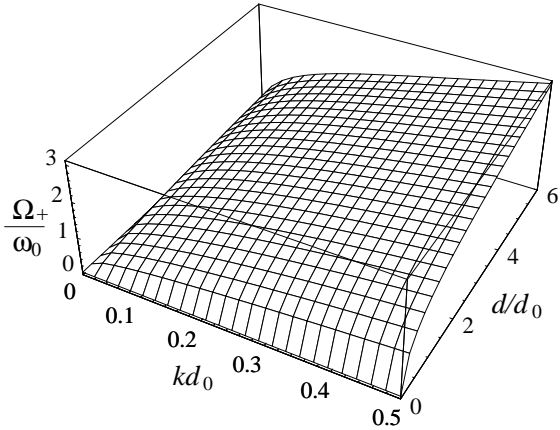}
\end{center}
\caption{Eigen-frequencies of the membrane correspond to in-phase oscillations of both layers together.}
\label{fig:finp}
\end{figure}

\begin{figure}[h]
\begin{center}
\includegraphics*[width=7.5cm]{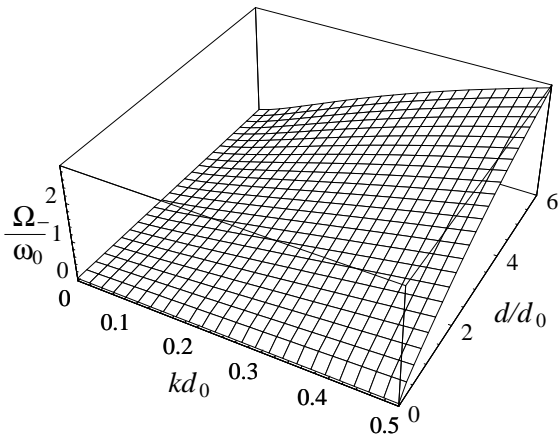}
\end{center}
\caption{Eigen-frequencies of the membrane correspond to out-of-phase oscillations of both layers together.}
\label{fig:fout}
\end{figure}

Finally, we shown in Figs.~\ref{fig:finp},~\ref{fig:fout} the dependence of the Eigen-frequencies on both the wave-number $k$ and membrane thickness $d$, where the thickness $d$ is given in units of $d_0$ and $k$ in units of $d_0^{-1}$. The frequency $\Omega_-(\mathbf k)$ corresponds to in-phase charge oscillations of the two layers and the frequency $\Omega_+(\mathbf k)$ to out-of-phase charge oscillations.

In order to obtain a more accurate expression of the plasmon frequencies instead of the bare mass and
surface density, one should use the effective mass of the hydrated ion and the effective nearly-free counterion
charge density. If there are binding sites on the membrane surface this will reduce the effective free
surface charge density.

The plasmon frequencies should scale as $\sqrt{\sigma}$ with respect to surface concentrations (which can be estimated from the bulk concentrations) and as $1/\sqrt{m^\ast_i}$ with respect to various counter-ion-effective masses. These plasmon oscillations can be used by the various parts of cell membrane to carry out communication. The investigations of the behavior of the gates of the voltage sensitive ion channels in cell membranes indicate a high sensitively of the voltage sensor to small changes of the membrane electric field. This implies that the $2D$ plasmas could play a role in triggering the bio-sensor operations. Before, however, one proceeds further to study such possibilities it is important to carry out an experimental study to verify
the existence and the properties of these plasmon modes.

\section{Conclusions}
\label{concl}

In this work we investigated the effects of the surface effective potential due to the existence of the charge carries at bounders of the bio systems, this problem can be simplified by considering the plasmon oscillations of each layer in cell membranes as a quasi-particle. We studied a theoretical model for plasmons modes of the cell membranes, by using Bogoliubov transformation method, we obtained the formulae of Eigen-frequencies of the membrane collective charge oscillations, the results are in good agreement with ones obtained from previous theories. This simple model could be applied for another bio, nano and soft material systems.

This work was supported by the National Fundamental Research Program on Natural Sciences of Vietnam, Physics Branch 4 029 06.

\end{document}